\documentclass[12pt]{article}
\usepackage{amsmath}
\usepackage{graphicx}
\usepackage{graphics}
\usepackage{amsmath}
\usepackage{rotating}
\usepackage{cite}
\usepackage{afterpage}
\usepackage{float}

\begin{document}

\date{October 25, 2000}
\title{\texttt{ New Method for Phase Transitions in Diblock Copolymers:}\\
-- The Lamellar Case}
\author{A. Rebei and J. De Pablo\thanks{%
depablo@engr.wisc.edu, rebei@cae.wisc.edu. } \\
%EndAName
Department of Chemical Engineering\\
University of Wisconsin - Madison.\\
Madison, WI 53706}
\maketitle

\begin{abstract}
A new mean-field type theory is proposed to study order-disorder transitions
(ODT) in block copolymers. The theory applies to both the weak segregation
(WSL) and the strong segregation (SSL) regimes. A new energy functional is
proposed without appealing to the random phase approximation (RPA). We find 
new terms unaccounted for within RPA. We work out 
in detail transitions to the lamellar state
and compare the method to other existing theories of ODT and numerical
simulations. We find good agreements with recent experimental results 
and predict that the intermediate segregation regime may have more 
than one scaling behavior.
\end{abstract}

\vspace{0.4in}
PACS numbers: 83.70.Hq, 36.20.-r, 61.25.Hq

\newpage

\section{Introduction}

\vspace{0in}

Predicting morphologies of block copolymers continues to be a very
challenging problem from the computational point of view. \hspace{0pt}To
make this latter task more feasible `nice' energy functionals are needed and
this is the goal of this paper. \hspace{0pt}For linear copolymers, the
self-consistent field method (SCFT) developed 
by Helfand and others \cite{helfand, hong, vasavour, shull, matsen, 
drolet, matsen2} is considered to be the method of choice. \hspace{0pt}It 
applies
quite well to all regimes of segregations. \hspace{0pt}Recently it was
successfully used to predict different morphologies of linear triblock
copolymers \cite{drolet}. \hspace{0pt}The method is based on a naive mean
field approximation to the partition \hspace{0pt}function. \hspace{0pt}%
Fluctuation effects can \hspace{0pt}also be added but at the expense of 
complicating the method 
\cite{shi}.

Around the order-disorder transition point, Leibler's \cite{leibler} field
theory for diblock copolymers is superior to the self-consistent field
theory. \hspace{0pt}For this reason there have been attempts to generalize
it to all types of segregations, intermediate as well as strong
segregations. \hspace{0pt}Ohta and Kawasaki \cite{ohta, kawasaki,
kawasaki2} were the first to propose a free energy functional of Leibler's
form that treats the strong segregation case of a diblock copolymer. It
gives comparable results as the self-consistent method and its predictions
are well supported by experiment \cite{papadakis, anastasiadis}. 
\hspace{0pt}Fluctuation effects were later added to Leibler's theory by
Fredrickson, Helfand and Barrat \cite{fredrickson,barrat,
brazovskii, mayes, muthukumar} who showed that the peak of the
scattering function not only depends on $\chi N$, the Flory-Huggins 
parameter, but also on the average segment length and
volume. 
\hspace{0pt}The Leibler free energy functional was also shown to be useful
in the strong segregation case if the wave vector dependence of the energy
functional is fully kept , since higher order spatial harmonics of the order
parameter become more and more important as the temperature is lowered in
the ordered phase\cite{melen, mcmullen}.

\vspace{0in}Here, we set to find a similar, but simpler, energy functional
for incompressible diblock copolymers that is valid for both the weak
 and strong segregation regimes. \hspace{0pt}Our work \hspace{%
0pt}has the same spirit as the recent work of Stepanow \cite{stepanow} where
he used graphical methods to obtain an improved self-consistent expression
for the structure factor of a diblock copolymer. \hspace{0pt}He avoided
using the random phase approximation (RPA)\cite{degennes} , i.e.,
Leibler's theory, and instead finds an expansion of the partition function
in terms of an effective potential \cite{edwards, vigis, 
weyersberg}. \hspace{0pt}Here we adopt the same goal of developing an
expression of the free energy based also on an effective potential. \hspace{%
0pt}The random phase approximation will not be used to get our energy
functional.

Besides succeeding in finding a new expression for the energy of a diblock
copolymer, we also find new ``ideal '' terms missed by RPA and already
found by Holyst and Vilgis \cite{holyst} in the polymer mixture case.

\vspace{0in}Our method is strictly functional; we do not use any graphs. 
\hspace{0pt}Our results to lowest order are similar to Stepanow result \cite
{stepanow} and our energy functional is qualitatively the same as Leibler's. 
\hspace{0pt}The quartic term is similar to Leibler's fourth order 
term but is much simpler to work with.

The paper is organized as follows. \hspace{0pt}In section II, we develop the
formalism. \hspace{0pt}The details are left to the Appendix. \hspace{0pt}In
section III, we solve the transition to the lamellar morphology of a
symmetric diblock. \hspace{0pt}In section IV, we compare our results with
those of the self-consistent method, Leibler's method, and simulations 
\cite{qy}.

\vspace{0in}

\section{Free Energy of a Block Copolymer}

We start by deriving a new expression for the free energy of a 
block copolymer melt. 
\hspace{0pt}Even though in this paper we are mainly interested in
incompressible diblock copolymers, it can be easily generalized to, e.g, 
triblocks. \hspace{0pt}The method we use  bypasses
 the random phase approximation and the
use of virtual sources, as was done originally by 
Leibler \cite{leibler}. \hspace{0pt}The Hamiltonian we use for 
our system is that of Edwards 
\cite{edwards} generalized to copolymers and can be taken of the form:

\vspace{0in}

\begin{align}
H& =H_{0}+V,  \notag \\
H_{0}[r_{i}]& =\frac{3}{2N\sigma ^{2}}\sum_{i=1}^{n}\int_{0}^{N}d\tau \text{ 
}\left( \frac{d\mathbf{r}_{i}(\tau )}{d\tau }\right) ^{2}  \notag \\
\mathbf{V}[r_{i}]& =\frac{1}{2}\int d\mathbf{r}d\mathbf{r}^{\prime
}\sum_{i,j=1}^{2}\rho _{i}(\mathbf{r})\rho _{j}(\mathbf{r}^{\prime })\mathbf{%
V}_{ij}(\mathbf{r},\mathbf{r}^{\prime }),  \notag \\
i,j& =A,B
\end{align}
$r_{i}$ is a curve along the i'th macromolecule that has two types of
monomers $A$ and $B$ with  densities $\rho _{A}(\mathbf{r})$ and $%
\rho _{B}(\mathbf{r})$, respectively. We assume both monomers to 
have the same Kuhn length 
$\sigma .$ There are $n$ chains in the melt, each with $\mathbf{f}$ $(%
\mathbf{f=}fN)$ monomers of type $A$ and $(N-\mathbf{f})$ of type $B.$ The
interaction potential $V$ is taken to have the simple form
 
\begin{equation}
\mathbf{V}(\mathbf{r},\mathbf{r}^{\prime })=\rho _{0}\left( 
\begin{array}{cc}
0 & \chi  \\ 
\chi  & 0
\end{array}
\right) \delta (r_{i}(s)-r_{j}(s^{\prime })),
\end{equation}
where $\chi $ is the Flory-Huggins constant and $\rho _{0}$ is the total
average density of monomers. The partition function of this incompressible
system of macromolecules is then given by 

\begin{equation}
\mathcal{Z}=\int d(r_{i})\delta (1-\rho _{A}(r)-\rho _{B}(r))\exp \left[
-\left( H_{0}+\frac{1}{2}\int d\mathbf{r}d\mathbf{r}^{\prime } \;
\overrightarrow{\mathbf{\rho }}^{T}(\mathbf{r})V(\mathbf{r},\mathbf{r}%
^{\prime })\overrightarrow{\rho }(\mathbf{r}^{\prime })]\right) \right] .
\end{equation}
In the above we have set the Boltzmann constant to one and used a vector
notation for the densities, i.e. 
\begin{equation}
\overrightarrow{\rho }(\mathbf{r})=\left( 
\begin{array}{c}
\rho _{A}(\mathbf{r}) \\ 
\rho _{B}(\mathbf{r})
\end{array}
\right) .
\end{equation}
The densities are given by 
\begin{equation}
\rho _{A}{(r)}=\frac{N}{\rho _{0}}\sum_{\alpha =1}^{n}\int_{0}^{f}ds \; 
\mathbf{\delta (r-r}_{\alpha }(s)) ,
\end{equation}
and 
\begin{equation}
\rho _{B}{(r)}=\frac{N}{\rho _{0}}\sum_{\alpha =1}^{n}\int_{f}^{1}ds \; 
\mathbf{\delta (r-r}_{\alpha }(s)) .
\end{equation}
In $\mathcal{Z}$, the second term in the exponential is a 
quadratic symmetric form and hence
it can be diagonalized. This diagonalization allows us to deal with a new
virtual set of monomers that are `decoupled'. \hspace{0pt}We therefore
introduce a new set of variables $\rho _{1}$ and $\rho _{2}$ such that: 
\begin{equation}
\rho _{1}(r)=\frac{1}{2}(\rho _{A}(r)+\rho _{B}(r)),
\end{equation}
\begin{equation}
\rho _{2}(r)=\frac{1}{2}(\rho _{A}{(r)}-\rho _{B}{(r)}) .
\end{equation}
In this new set of variables the potential becomes diagonal with 
\begin{equation}
V_{11}=2\rho _{0}\chi ,
\end{equation}
\begin{equation}
V_{22}=-2\chi \rho _{0} ,
\end{equation}
\begin{equation}
V_{12}=V_{21}=0 .
\end{equation}
Ignoring the incompressibility condition for now and after isolating the
free energy of the disordered state we have 
\begin{align}
\mathcal{Z}& =\int \mathcal{D}r_{i}\exp \{-\frac{1}{2}\Delta \rho _{\alpha }%
\mathbf{V}_{\alpha \beta }\Delta \rho _{\beta }-\rho _{\alpha }^{0}\mathbf{V}%
_{\alpha \beta }\Delta \rho _{\beta }\}  \notag \\
& \times \exp \{-\frac{1}{2}\rho _{\alpha }^{0}\mathbf{V}_{\alpha \beta
}\Delta \rho _{\beta }^{0}\} ,
\end{align}
where 
\begin{equation*}
\mathcal{D}r_{i}=d(r_{i})\exp (-H_{0}),
\end{equation*}
and 
\begin{align*}
\Delta \rho _{\alpha }(\mathbf{r})  =\rho _{\alpha }(\mathbf{r})-\rho
_{\alpha }^{0} , \; \; \; \;  \alpha  = 1,2 .
\end{align*}
For a symmetric diblock, 
\begin{equation}
\rho ^{0}=\binom{\frac{1}{2}}{0} .
\end{equation}
Upon introducing a two-component Hartree-type field $\varphi_{\alpha}$
 with which these
new virtual molecules are interacting,  the partition function becomes 
\begin{equation}
\begin{array}{cc}
\mathcal{Z}= & \int \mathcal{D}(r_{i})\mathcal{D}\mathbf{\varphi }\exp [-%
\frac{1}{2}\int d\mathbf{r}d\mathbf{r}^{\prime }\mathbf{\varphi }_{\alpha }(%
\mathbf{r})\mathbf{V}_{\alpha \beta }^{-1}(\mathbf{r},\mathbf{r}^{\prime })%
\mathbf{\varphi }_{\beta }(\mathbf{r}^{\prime }) \\ 
& +i\int d\mathbf{r[\varphi }_{\alpha }(\mathbf{r})+i\rho _{\beta }^{0}%
\mathbf{V}_{\alpha \beta }]\cdot \Delta \mathbf{\rho }_{\alpha }(\mathbf{r}%
)].
\end{array}
\end{equation}
We introduce now two more fields, $\Phi $ and $\mu $. \hspace{0pt}Since 
\begin{equation}
\int \mathcal{D\Phi }_{\alpha }\; \delta (\Phi _{\alpha }(\mathbf{r})-\Delta
\rho _{\alpha }(\mathbf{r}))=1,
\end{equation}
then we have 
\begin{align}
\mathcal{Z}& =\exp (-F_{0})\int \mathcal{D}\varphi \mathcal{D}\Phi \mathcal{%
D\mu D(}r_{i})\exp \{-i\mu _{\alpha }\Phi _{\alpha }\}  \notag \\
& \times \exp \{-\frac{1}{2}\varphi _{\alpha }\mathbf{V}_{\alpha \beta
}^{-1}\rho _{\beta }+i(\rho _{\alpha }+\mu _{\alpha }+i\rho _{\beta }^{0}%
\mathbf{V}_{\alpha \beta })\Delta \rho _{\alpha }\} ,
\end{align}
where 
\begin{equation}
F_{0}=n\chi Nf(1-f) .
\end{equation}
is the energy of the disordered state. Now, we make a change of variables
and let

\vspace{0in}

\begin{equation}
\overline{\varphi }_{\alpha }(r)=\varphi _{\alpha }(r)+\mu _{\alpha
}(r)+i\int dr \; V_{\alpha \beta }(r)\rho _{\beta }^{0}(r) .
\end{equation}
We also define a new functional 
\begin{equation}
\mathbf{F}[\varphi ]\equiv \ln \left( \int \mathcal{D}(r_{i})\exp \left[
-\left( \mathbf{H}_{0}(r_{i})+i\int d\mathbf{r}\; 
\overline{\mathbf{\varphi }}(%
\mathbf{r})\cdot \mathbf{\rho }(\mathbf{r})\right) \right] \right) .
\end{equation}
This functional can be expanded in $\overline{\varphi }$ around a homogeneous
state

\begin{eqnarray}
\mathbf{F}[\varphi ]& =\sum_{m=0}^{\infty }\frac{1}{m!}\sum_{\left\{ \alpha
\right\} }\int \int ...\int d\mathbf{r}_{1}d\mathbf{r}_{2}...d\mathbf{r}_{m}%
{\ C}_{_{\alpha 1\alpha 2}...\alpha m}\left( \mathbf{r}_{1},\mathbf{r}%
_{2},...,\mathbf{r}_{m}\right)  \\
& \times \varphi _{\alpha _{1}}(\mathbf{r}_{1})\varphi _{\alpha _{2}}(%
\mathbf{r}_{2})...\varphi _{\alpha _{m}}(\mathbf{r}_{m})  \notag \\
& \alpha _{i} =1,2.  \notag
\end{eqnarray}
The coefficients $C_{\alpha \beta ...}$ are given in Appendix A. In the 
following we keep only terms up to the fourth order.
After some rearrangements and writing $\varphi $ in place of $\overline{%
\varphi }$, we have the following expression for the partition function,
\begin{align}
\mathcal{Z}& =\int \mathcal{D\Phi D}{\mu }\mathcal{D\varphi }\exp \{-\frac{1%
}{2}\varphi _{\alpha }(V_{\alpha \beta }^{-1}+C_{\alpha \beta })\varphi
_{\beta } \\
& -\frac{1}{4!}C_{\alpha \beta \lambda \gamma }\varphi _{\alpha }\varphi
_{\beta }\varphi _{\lambda }\varphi _{\gamma }+i\Xi _{\alpha }\varphi
_{\alpha }\}  \notag \\
& \times \int \mathcal{D\Psi }\exp \{\frac{1}{4!}C_{\alpha \beta \lambda
\gamma }(\Psi _{\alpha \beta }-\varphi _{\alpha }\varphi _{\beta })(\Psi
_{\lambda \gamma }-\varphi _{\lambda }\varphi _{\gamma })\} .  \notag
\end{align}
Here we have introduced a new pairing field $\Psi _{\alpha \beta }$
so we can cancel the  quartic term in $\varphi _{\alpha }$. \hspace{0pt}%
We have also set 
\begin{equation}
i\Xi (\mathbf{r})=i\rho _{\alpha }^{0}+V_{\alpha \beta }^{-1}\mu _{\beta }(%
\mathbf{r}),
\end{equation}
and have not written the space integrals explicitly.
Then, we can integrate the $\varphi $ field and then the $\mu $ \hspace{0pt}%
field exactly. \hspace{0pt}We are left with only two fields $\Phi $ and $%
\Psi $ ( see Appendix B) 
\begin{equation}
\mathcal{Z}=\int \mathcal{D\Phi D}\Psi \exp \{-\mathcal{F(}\Phi ,\Psi )\} ,
\end{equation}
where 
\begin{align}
\mathcal{F(}\Phi ,\Psi )=-\frac{1}{4!}\Psi _{\alpha \beta }C_{\alpha \beta
\lambda \gamma }\Psi _{\lambda \gamma }& +\frac{1}{2}\rho _{\alpha
}^{0}A_{\alpha \beta }^{-1}\rho _{\beta }^{0}  \notag \\
+\frac{1}{2}\zeta _{\alpha }^{0}B_{\alpha \beta }^{-1}\zeta _{\beta }^{0}+%
\frac{1}{2}\log \det A_{\alpha \beta }& +\frac{1}{2}\log \det B_{\alpha
\beta },
\end{align}
and 
\begin{align}
\zeta _{\alpha }(\mathbf{r})& =\Phi _{\alpha }(\mathbf{r})+\rho _{\alpha
}^{0}-V_{\alpha \beta }^{-1}A_{\beta \lambda }^{-1}\rho _{\lambda }^{0}, \\
A& =(1+\frac{1}{6}\Delta U)U^{-1} , \notag \\
B& =V^{-1}-V^{-1}A^{-1}V , \notag \\
\Delta _{\alpha \beta }& =\Psi _{\lambda \gamma }C_{\lambda \gamma \alpha
\beta } , \notag \\
U_{\alpha \beta }& =(V_{\alpha \beta }+C_{\alpha \beta })^{-1}.  \notag
\end{align}
In the above $A,$ $B,$ $\Delta ,$ and $U$ are spatial 
dependent matrices. $U_{\alpha \beta }$
is the desired effective potential \cite{edwards}. \hspace{0pt}We 
expand the logarithmic
terms in powers of $U$. \hspace{0pt}We do the same when we seek an
expression for $B.$ \hspace{0pt}In all expansions, we keep only quadratic
terms in $\Psi .$ \hspace{0pt}The final lowest order expression we get from
this expansion is our expression for the energy

\begin{eqnarray}
H(\Phi ) & = & \frac{1}{2 vol} \sum_{q} \Phi (q) ( \frac{1}{C_{22}(q)} - 
2\chi \rho _{0}  ) \Phi (-q)  \\
& & -\frac{1}{3!}\frac{1}{(vol)^{2}}\sum_{p} \Phi (q)  
\frac{C_{2222}(q,-q,p,-p)}{
[C_{22}(q)]^{2}C_{22}(p)} \Phi (-q)  \nonumber \\
& &-\frac{1}{4!}\frac{1}{(vol)^{3}}\sum_{q_{1}q_{2}q_{3}}\frac{%
C_{2222}(q_{1},q_{2},q_{3,}-q_{1}-q_{2}-q_{3})}{%
C_{22}(q_{1})C_{22}(q_{2})C_{22}(q_{3})C_{22}(q_{1}+q_{2}+q_{3})}  \nonumber \\
& &  \times \Phi (q_{1})\Phi (q_{2})\Phi (q_{3})\Phi (q_{1}+q_{2}+q_{3}). 
\nonumber
\end{eqnarray}

It can be easily shown that $vol/(2 n C_{22}(q))$ is actually Leibler's
structure function. Leibler's structure function $S(q)$ is given by
(Appendix A) 
\begin{equation}
S^{-1}(q)=\frac{S_{AA}(q)+2 S_{AB}(q)+S_{BB}(q)}{%
S_{AA}(q)S_{BB}(q)-S_{AB}^{2}(q)}.
\end{equation}
This is indeed equivalent to $C_{22}^{-1}(q)$ with the proper normalization.
However it is easier to see this graphically and in fig$.1$, we plot this
function. \hspace{0pt}Hence the first and the second terms are the usual $RPA
$ result for the inverse scattering function. \hspace{0pt}The third term is
new and is not captured by the $RPA$ approximation. \hspace{0pt}It is not
due to fluctuations. \hspace{0pt}It has been first pointed out by Holyst and
Vilgis in their study of polymer blends \cite{holyst}. \hspace{0pt}The
fourth term is familiar but the coefficient is much  simpler to
calculate and behaves differently than Leibler's term for large wave
vectors. In fig.2, we plot both terms for a subset of wavenumbers. Our fourth
order coefficient is given by

\begin{figure}[H]
\begin{turn}{-90}
\resizebox{\textwidth}{!}
{\includegraphics[0in,0in][5.5in,8in]{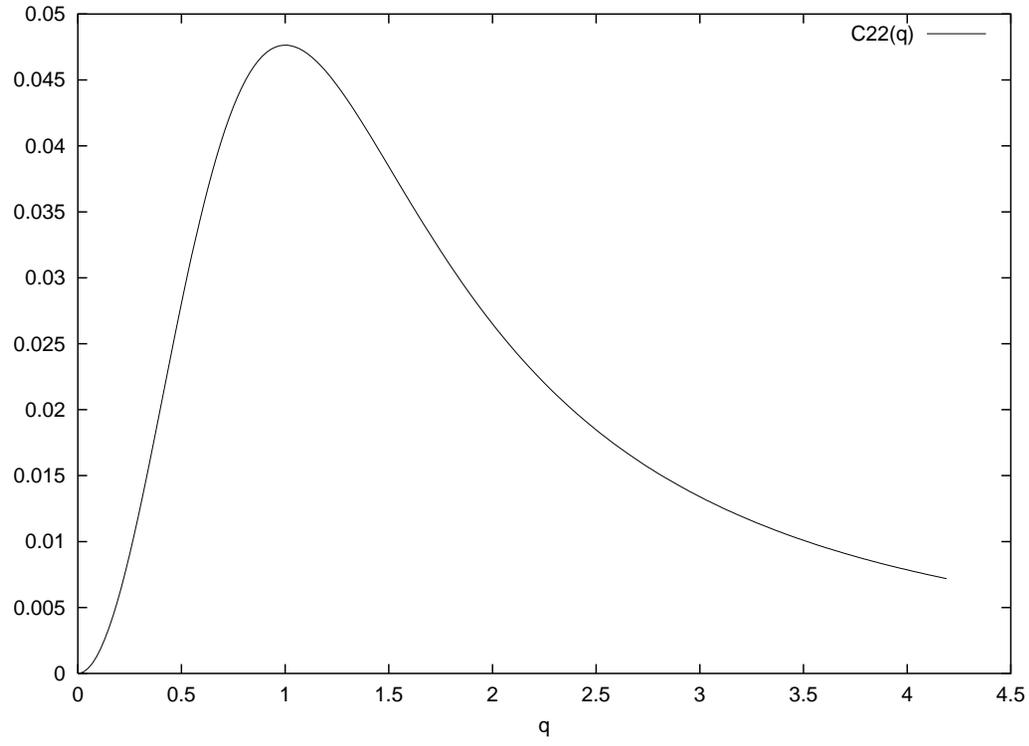} }
\end{turn}
\caption{The two body correlation function, $C_{22}(q)$, in
our diagonalized system of collective densities. }
\end{figure}

\begin{figure}[H]
\resizebox{\textwidth}{!}
{\includegraphics[0in,0in][8in,10in]{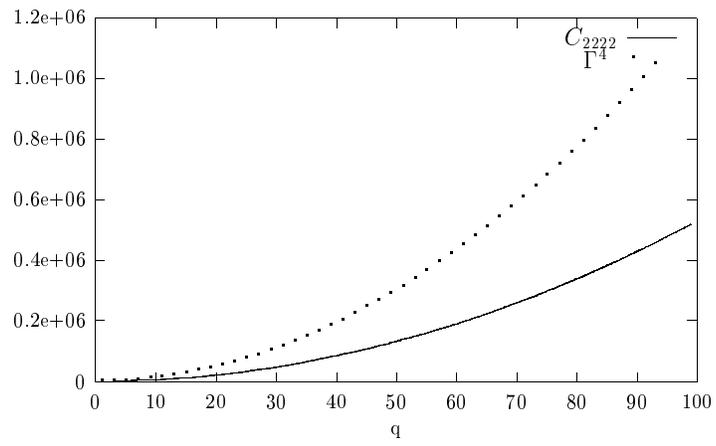}}
\caption{ Comparison of fourth order coefficients of this theory
and Leibler's RPA theory. }
\end{figure}

\begin{equation}
C^{(4)}(q_{1},q_{2},q_{3,}-q_{1}-q_{2}-q_{3})=\frac{%
C_{2222}(q_{1},q_{2},q_{3,}-q_{1}-q_{2}-q_{3})}{%
C_{22}(q_{1})C_{22}(q_{2})C_{22}(q_{3})C_{22}(q_{1}+q_{2}+q_{3})}.
\end{equation}
This term is independent of any three-body correlation functions. Leibler's $%
\Gamma ^{(4)}$ term does however depend on three body correlation terms, and
this greatly complicates  computations involving it; it is given by \cite{leibler} 
\begin{equation}
\begin{array}{cc}
\Gamma ^{(4)}(q_{1},q_{2},q_{3,}-q_{1}-q_{2}-q_{3}) & =\gamma
_{ijkl}(q_{1},q_{2},q_{3,}-q_{1}-q_{2}-q_{3})(S_{iA}^{-1}(q_{1}) \\ 
& -S_{iB}^{-1}(q_{1}))(S_{jA}^{-1}(q_{2})-S_{jB}^{-1}(q_{2}))\times
(S_{kA}^{-1}(q_{3}) \\ 
& -S_{kB}^{-1}(q_{3}))(S_{lA}^{-1}(q_{4})-S_{lB}^{-1}(q_{4})).
\end{array}
\end{equation}
with $i,j,k,l=A,B$. $\gamma _{ijkl}$ is a function of two-point, three-point
and four-point correlation functions. In fig.2, we plot both coefficients
for $q_{1}=-q_{2}=q_{3}=-q_{4}$.

To lowest order, the  propagator of this theory also agrees with that of Stepanow \cite
{stepanow}, except that our self-energy term
has additional contributions from the fourth term, $Tr(\Lambda U)$, 
of the free energy in Eq.(B-7). \hspace{0pt}Hence to consider any
fluctuations this term must be included from the outset and therefore the
work in \cite{stepanow} is more thorough than the one presented, e.g., in 
\cite{barrat} and others based on RPA. In the next section where we use 
our functional to study
transitions from a disordered state to a lamellar state, we will not 
include the quadratic non-RPA term in
order to compare our
functional with that of Leibler's. 

\section{The Lamellar Solution}

\vspace{0in}

\vspace{0in}Following Melenkevitz and Muthukumar\cite{melen}, we conjecture
a solution that minimizes the energy functional. Knowing that in SCFT 
the densities are found by solving a modified heat equation, we choose a
function of the following form 
\begin{equation}
\Phi (x)=\sum_{l=1,3,...}\frac{2}{\pi l}\exp [-\frac{1}{2}(q_{l}\lambda
)^{2}]\sin (q_{l}x) .
\end{equation}
This choice is also dictated by the fact that 
\begin{equation}
\int dx\Phi (x)=0,
\end{equation}
and the solution must be periodic. \hspace{0pt}The wave vector $q$ is given
by 
\begin{equation}
q_{l}=\frac{2\pi l}{D},
\end{equation}
where $D$ is the lamellar periodicity. \hspace{0pt}$\lambda $ is another
parameter besides \hspace{0pt}$D$ that is related to the wall thickness
of the interface region between the two components 
of the diblock copolymer.  \hspace{0pt}Both parameters are to be found by
minimizing the energy $H$ with respect to them. \hspace{0pt}Actually, we
solve for $\lambda $ for a given $D$ and then calculate the corresponding
energy and choose the solution with the lowest energy. \hspace{0pt}We
rescale dimensions in terms of the radius of gyration so the energy per
chain is given by 
\begin{align}
H/n& =\frac{1}{2}\sum_{m=1,3,..}b_{m}^{2}\left( \frac{vol}{2nC_{22}(m)}-\chi
N\right)  \\
& -\frac{1}{384}\sum_{\substack{ m=\pm 1,\pm 3... \\ p \\ r \\ s=-m-p-r}}%
\frac{b_{m}b_{p}b_{r}b_{s}C_{2222}(m,p,r,s)}{%
C_{22}(m)C_{22}(p)C_{22}(r)C_{22}(s)},  \notag
\end{align}

\vspace{0in}where 
\begin{equation}
b_{m}=\frac{2}{\pi m}\exp\lbrack-\frac{1}{2}(2\pi m\lambda/D)^{2}\rbrack .
\end{equation}

Figures 3-8 summarize all the results about this particular solution. In 
particular, we observe that this solution predicts that the order-disorder
transition (ODT) occurs for $\chi N $ right below $10.5$. Immediately below
the transition temperature, our energy functional shows that the behavior of
the polymer chains is no longer Gaussian. For Gaussian chains, the scaling
factor $\delta$ between $\ln(D)$ and $\ln(\chi N)$ is zero. In our case, we 
find that for $\chi N$ less than 13, $\delta$ is approximately equal to
0.26. For $\chi N$ between 13 and 30, $\delta = 0.52$, and for 
$\chi N $ above 30, $\delta$ becomes about 0.19. Hence the $2/3$-power law
between $D$ and $N$ in the SSL is also
verified by this solution. So far only one intermediate region has 
been observed \cite{papadakis}. Here, this solution suggests that the 
intermediate region is really more than one. Since the behavior below the
 ODT is believed to be nonuniversal, it will be interesting to see if this 
predicted behavior is also observed in some symmetric diblock copolymers 
other than the one treated in \cite{papadakis}. The 
inclusion of the non-RPA term that we omitted in this solution will
not change this overall picture. It has 
only a moderate effect at high values of 
$\chi N$ where higher and higher wavenumbers are needed in the energy.      
\begin{figure}[H]
\resizebox{\textwidth}{!}
{\includegraphics[0in,0in][8in,10in]{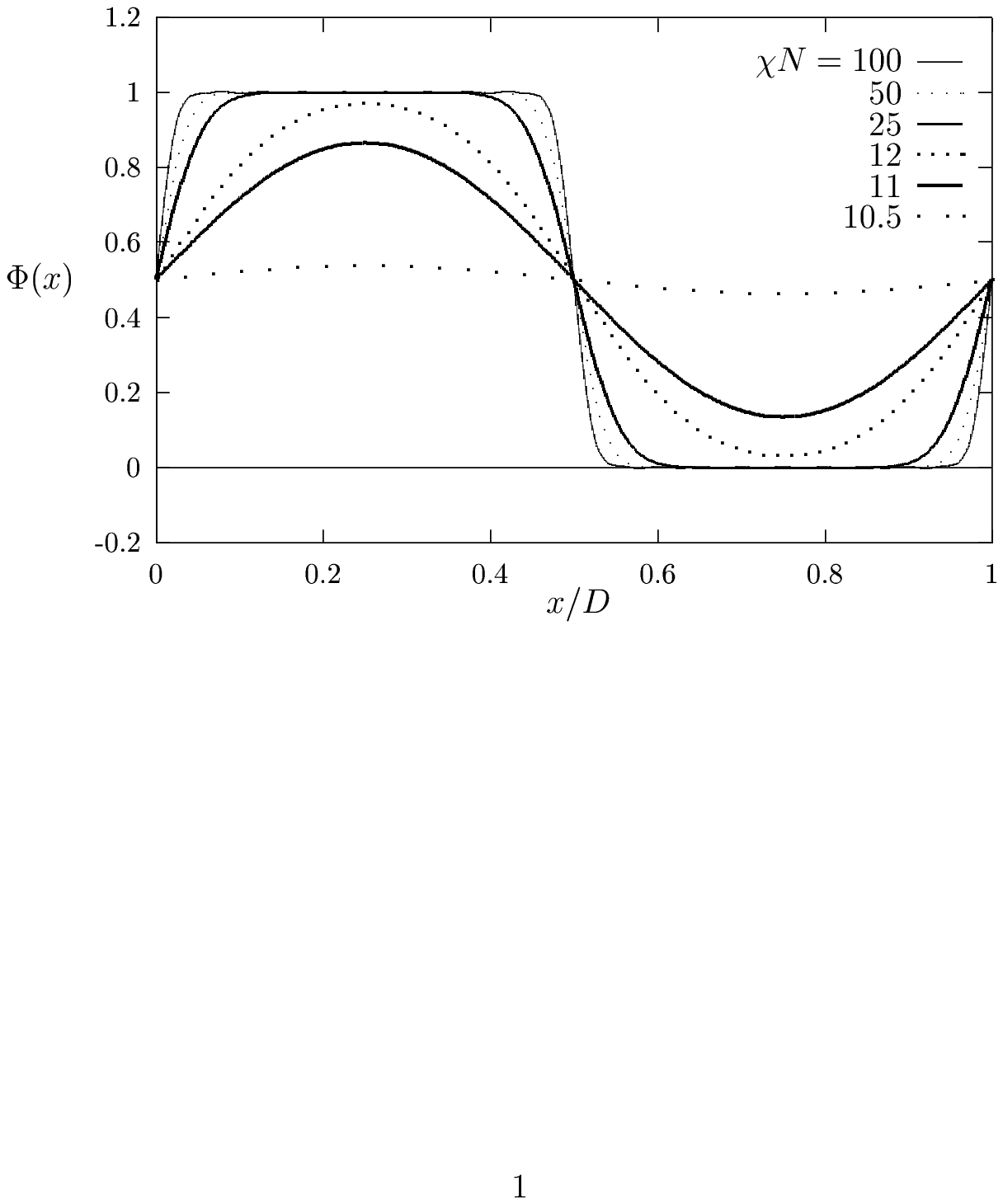}}
\caption{ Density profiles for the lamellar morphology given by
Eq.(30) for $\chi N = 10.5, 11, 12, 25, 50, 100$. }
\end{figure}

\vspace{0in}

\section{\protect\vspace{0in}Comparison and discussion:}

In fig.3, we plot  density profiles for different $\chi N$'s. We
observe that the order-disorder transition occurs below $\chi N=10.5$. The
segregation amplitude grows much faster for $\chi N$ just above the ODT
temperature $\chi N_{c}$ and less than 30. Hence in this theory, the strong
segregation regime is attained much faster than the self-consistent method
(SCFT) predicts \cite{matsen}. This is also confirmed by recent experiements
on symmetric diblock copolymers \cite{papadakis}. Figure 4 compares our results
to that of the SCFT calculations at $\chi N=12.5$ and simulations \cite{qy}.
The simulation was done with chains of forty-eight segments. Assuming that
the critical temperature, $\chi N_{c}$, for these chains is also close to
10.495 as is the case for infinite chains, we find that the simulation gives
a result that falls between our result and the SCFT result. However the ODT
temperature for these short chains is expected to be less than in the ideal
case. The SCFT curves were found by solving Eqs.(3-7), that appear in 
\cite{matsen}, using a finite difference method and using a random 
configuration as an input. Our results agree well with those given in 
\cite{matsen2}. In fig.5, we instead predict the $\chi N$ value for which our theory
coincides with the simulation results. We find a $\chi N$ value of
approximately 11.15. Assuming now that $\chi N /{\chi N_{c}}=1.2$
corresponds to $\chi N=11.15$ in our theory, we plot in fig.6 and fig.7 density
profiles for $\chi N=22.3$ and $\chi N=44.5$ and compare our results against
those of simulation. We find relatively good agreements between theory and
simulation especially far from the interfaces. Finally, in fig.8, we check
if our theory predicts the observed scaling behavior. Clearly, we can
distinguish three regimes from the plot. The intermediate regime extends
 from about $\chi N=13$ and  to about $\chi N=27$, has a scaling
factor $\delta$ of about 0.52. Above $\chi N=30$, the scaling factor 
is about 0.19,
in agreement with observations \cite{papadakis}. Compared to Leibler's
energy functional \cite{mcmullen}, our results are much closer to the SCF.
This is due to the fourth order term which is smaller than Leibler's
fourth order term for large wavenumbers. For large $\chi N$, our energy
functional gives closer periods to the SCFT method than the full
Leibler Hamiltonian \cite{mcmullen} that overestimates periods by as
much as 30 \%. Moreover, our theory predicts that transition to the 
strong segregation regime occurs at $\chi N \approx 30$. In complete 
agreement with the experimental results in \cite{papadakis}. In the SCFT, 
the SSL is believed to occur around $\chi N = 50$. Ref. \cite{melen} 
suggests that
the SSL starts to occur for $\chi N$ larger than $90$. 

\begin{figure}[H]
\resizebox{\textwidth}{!}
{\includegraphics[0in,0in][8in,10in]{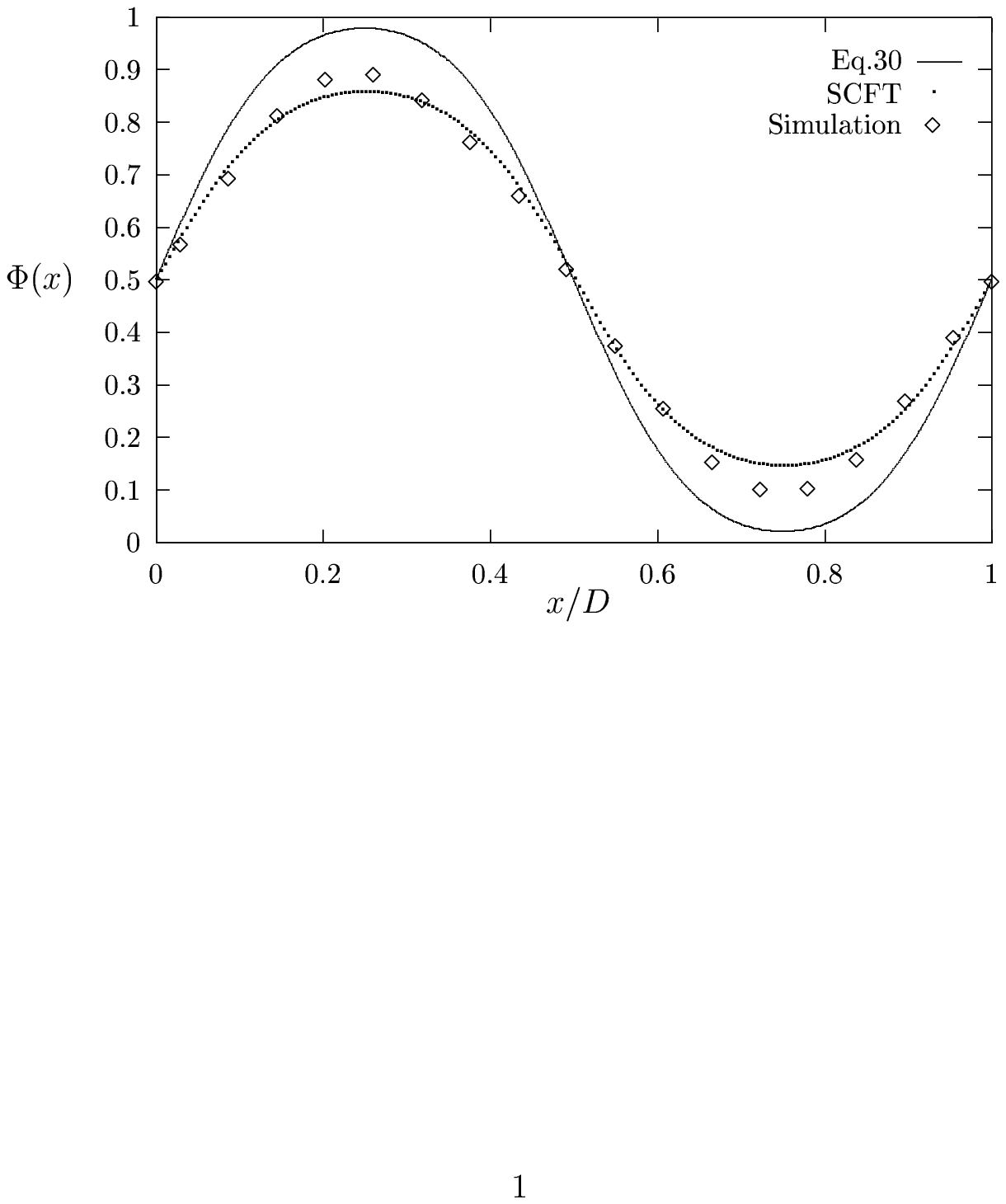}}
\caption{Density profiles of the symmetric lamellar solution for $%
\chi N = 12.5$. The Simulation is done for ${\chi N}/{\chi N_{c}} = 1.2$. }
\end{figure}

\begin{figure}[H]
\resizebox{\textwidth}{!}
{\includegraphics[0in,0in][8in,10in]{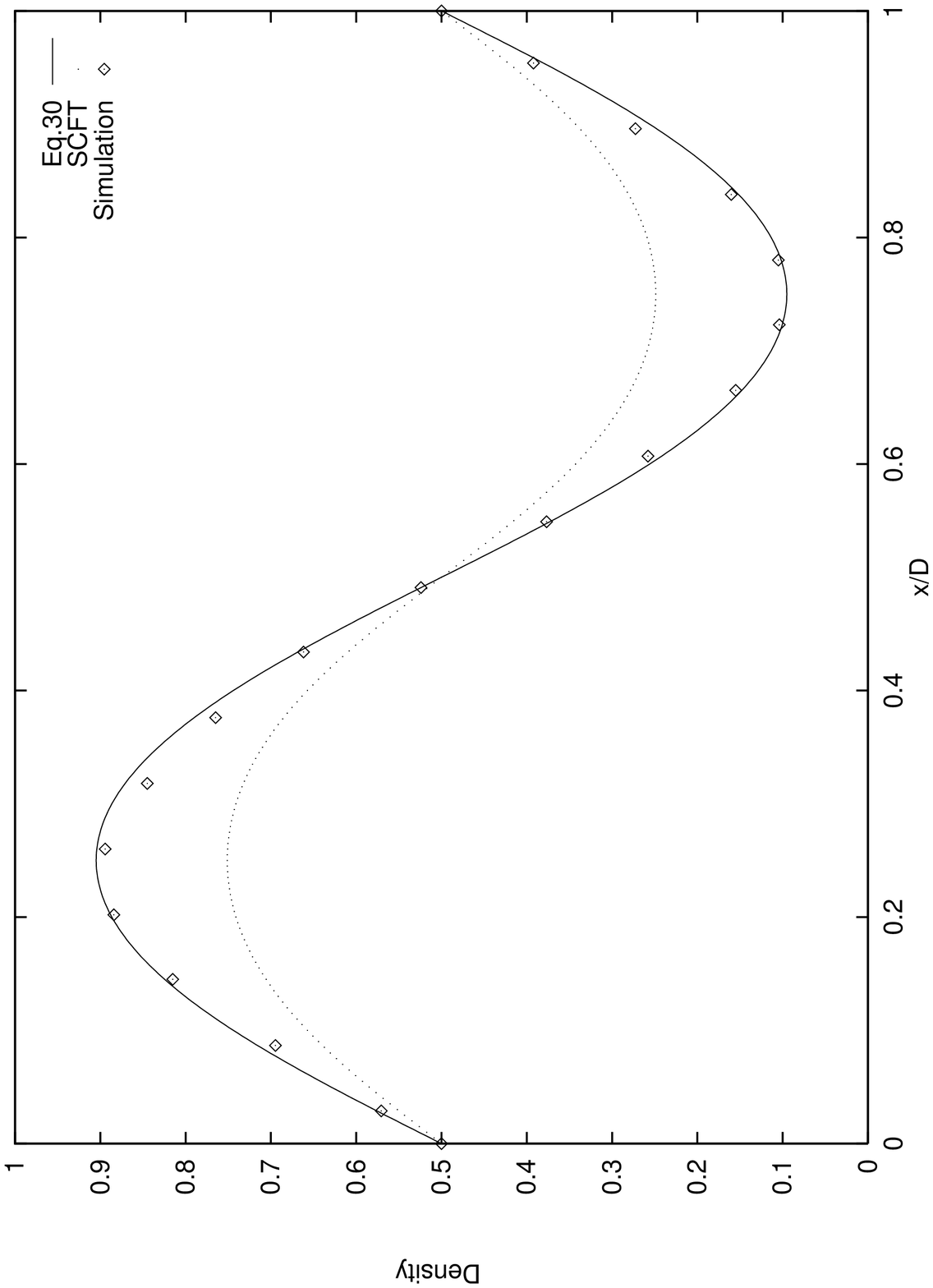}}
\caption{ Comparison of density profiles for $\chi N = 11.15$. The
simulation curve is at ${\chi N}/{\chi N_{c}} = 1.2$ }
\label{c11}
\end{figure}

\begin{figure}[H]
\resizebox{\textwidth}{!}
{\includegraphics[0in,0in][8in,10in]{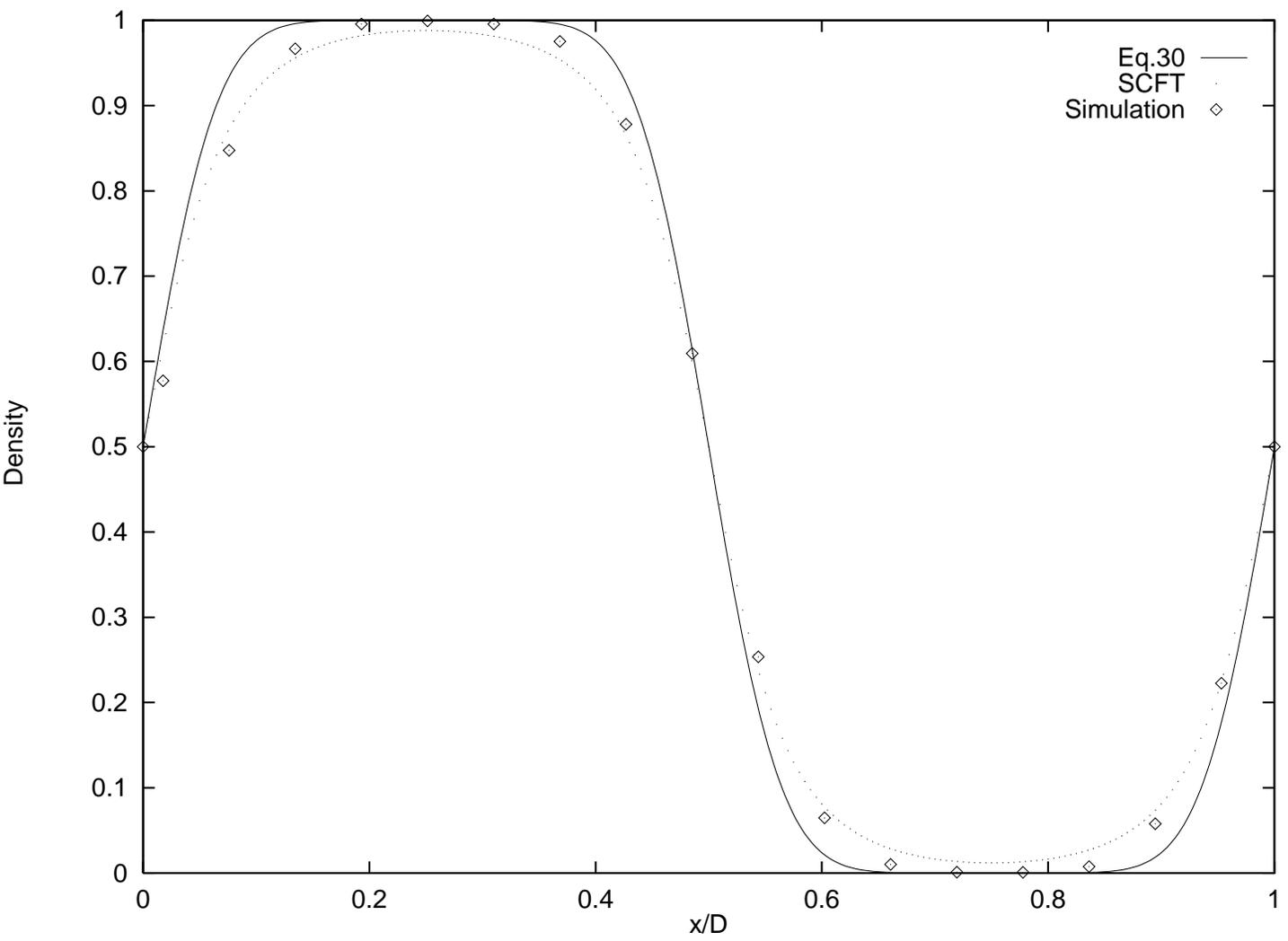}}
\caption{Comparison of density profiles for $\chi N = 22.3$. The
simulation curve is at ${\chi N}/{\chi N_{c}} = 2.4$. }
\label{c22}
\end{figure}

\begin{figure}[H]
\resizebox{\textwidth}{!}
{\includegraphics[0in,0in][8in,10in]{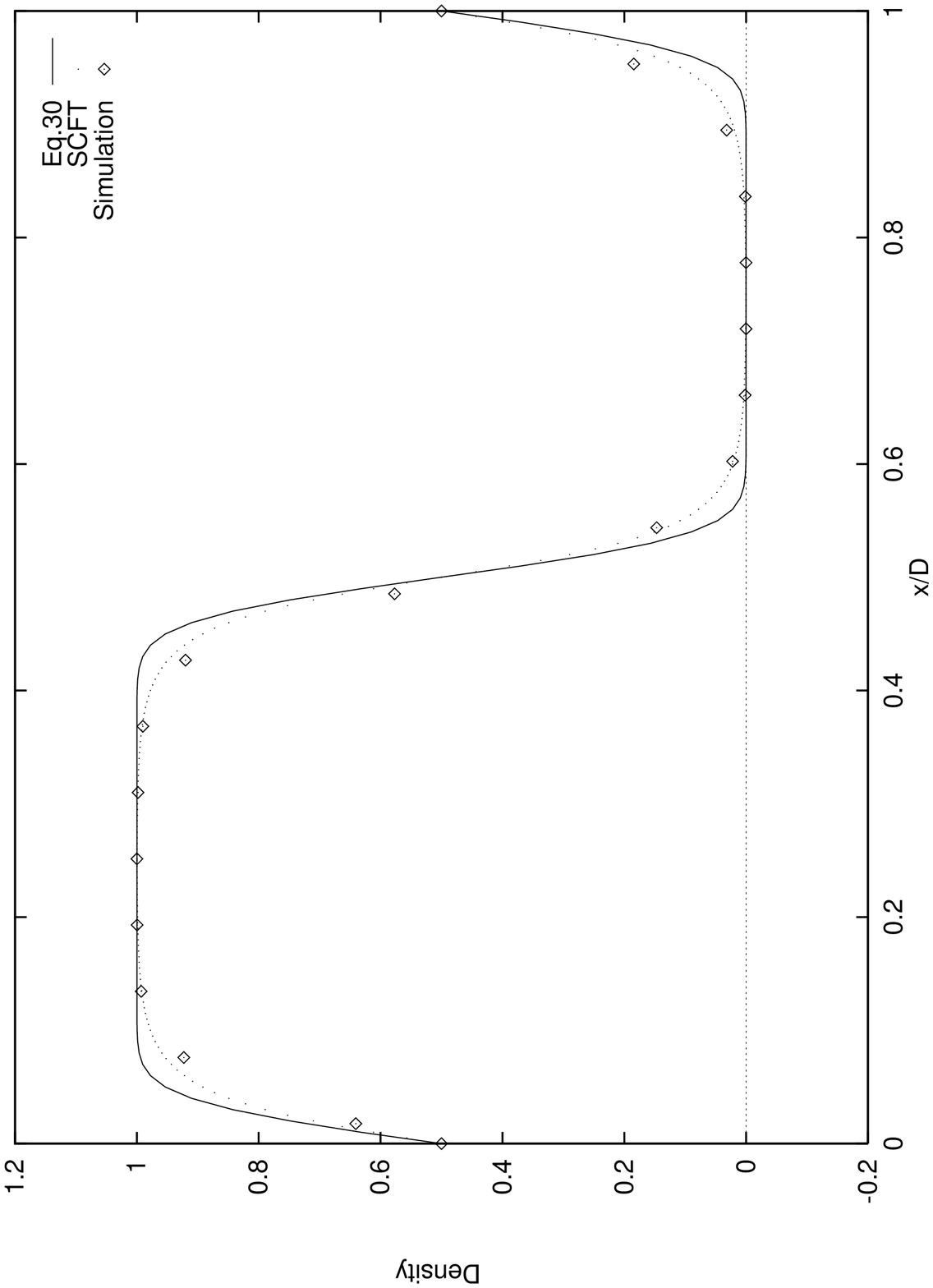}}
\caption{Comparison of density profiles for $\chi N = 44.5$. The
simulation curve is at ${\chi N}/{\chi N_{c}} = 4.8$. }
\label{c44}
\end{figure}

\vspace{1.0in}
\begin{figure}[H]
\resizebox{\textwidth}{!}
{\includegraphics[0in,0in][8in,10in]{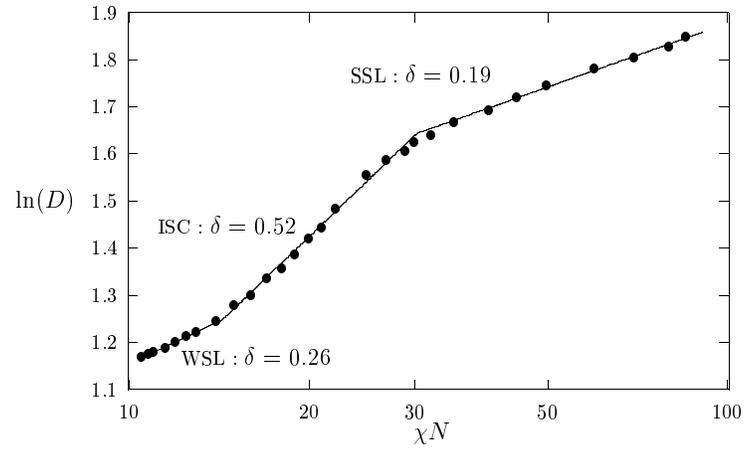}}
\caption{lot of $\ln D$ vs. $\ln(\chi N)$ for a symmetric
lamellar morphology. For high $\chi N$, the scaling factor is approximately
0.69. $\delta$ is the slope of the best fit segment to the data in three
different regions. The weak segregation limit (WSL), the intermediate 
segregation configuration (ISC) and the strong segregation limit (SSL). }
\end{figure}

\vspace{0.5in} 

\section*{Acknowledgments}

We  thank Dr. Qiliang Yan for 
providing us with the simulation results. 
A.R. also thanks Prof. C. Goebel for many stimulating discussions and for his
comments on the manuscript.

\newpage

\section*{Appendix}

\subsection*{ Appendix A}

Let $Q$ be a partition function of a melt of polymer chains in  external 
fields  $\varphi_{1}$ and  $\varphi_{1}$ 
\begin{equation}
Q[\varphi _{i}]=\int \mathcal{D}(r)\exp \{-i\varphi _{1}\Sigma \phi
-i\varphi _{2}\Delta \phi \}  \tag{A-1}
\end{equation}
where
\begin{align}
\Delta \phi & =(1-f)\varphi _{1}-f\varphi _{2}  \tag{A-2} \\
\Sigma \phi & =\varphi _{1}+\varphi _{2}-1  \notag
\end{align}
 $\varphi _{1}$, $\varphi _{2}$ are densities.

We next expand $\log Q$ in terms of $\varphi_{1}$ and  $\varphi_{2}$ . Since 
$Q$ $\hspace{0pt}$%
should be invariant under $\Delta\phi\longrightarrow-\Delta\phi$ and $%
\Sigma\phi\longrightarrow-\Sigma\phi,$ only even powers of $\varphi_{i}$ are
present, hence we write: 
\begin{align}
\log Q & =\frac{-1}{2!}\int dx_{1}dx_{2} \; C_{\alpha\beta}(x_{1},x_{2})
\varphi_{\alpha
}(x)\varphi_{\beta}(y)  \tag{A-3} \\
& +\frac{1}{4!}\int
dx_{1}...dx_{4}C_{\alpha\beta\gamma\delta}(x_{1},...,x_{4})\varphi_{%
\alpha}(x_{1})...\varphi_{\delta}(x_{4})  \notag \\
& +...  \notag
\end{align}
where, e.g., 
\begin{equation}
C_{22}(x_{1},x_{2})=<\Delta \varphi (x_{1})\Delta \varphi (x_{2})>_{0}  \tag{A-4}
\end{equation}
\begin{align}
C_{2222}(x_{1}...x_{4})& =<\Delta \varphi (x_{1})...\Delta \varphi
(x_{4})>_{0}  \tag{A-5} \\
& -C_{22}(x_{1},x_{2})C_{22}(x_{3},x_{4})  \notag \\
& -C_{22}(x_{1},x_{3})C_{22}(x_{2},x_{4})  \notag \\
& -C_{22}(x_{1},x_{4})C_{22}(x_{2},x_{4})  \notag
\end{align}

\vspace{0.3in}

The averages $<...>_{0}$ are evaluated with a Gaussian distribution. \hspace{%
0pt}Hence we find 
\begin{equation*}
C_{22}(q)=S_{AA}(q)-2S_{AB}(q)+S_{BB}(q)
\end{equation*}
with 
\begin{align}
S_{AA}(x)& =\frac{2}{x^{2}}(f_{A}(x)+\exp (-f_{A}(x)-1)  \tag{A-6} \\
S_{AB}(x)& =\frac{-1.0}{x^{2}}(\exp (-f_{B}(x))-1-\exp (-x)+\exp (-f_{A}(x))
\notag \\
\text{and \hspace{0pt}}x& =q^{2}\frac{\sigma ^{2}N}{6}  \notag
\end{align}
Similar expressions for the other correlation functions, such 
as $C_{2222}$, can straightforwardly
derived, but with much more labor. It amounts to calculating all 
correlation functions, $G_{\alpha \beta \gamma \delta}(x_{1},
x_{2},x_{3},x_{4})$, in the Gaussian distribution of the form

\begin{align}
G_{\alpha \beta \gamma \delta}(x_{1},x_{2},x_{3},x_{4}) & = 
\int ds_{1,\alpha} \int ds_{2,\beta} \int ds_{3,\gamma} \int ds_{4,\delta}
\langle \delta(x_{1} - x(s_{1,\alpha}))\delta(x_{2} -   \tag{A-7} \\
&  x(s_{2,\beta}))\delta(x_{3} - x(s_{3,\gamma}))
\delta(x_{4} - x(s_{4,\delta})) \rangle_{0}  \notag
\end{align}
where $\alpha, \beta, \gamma, \delta = 1,2 $. Hence $C_{2222}$ will be 
 a linear combination of all these functions.

\subsection*{ Appendix B}

We start from 
\begin{align}
\mathcal{Z} & =\int\mathcal{D\Phi D}\Psi\exp\{-i\mu_{\alpha}\Phi_{\alpha }-%
\frac{1}{2}\mu_{\alpha}V_{\alpha\beta}^{-1}\mu_{\beta}-i\rho_{\alpha}^{0}%
\mu_{\alpha}\}  \tag{B-1} \\
& \times\int\mathcal{D\varphi}\exp\{-\frac{1}{2}\varphi_{\alpha}(V_{\alpha%
\beta}^{-1}+C_{\alpha\beta})\varphi_{\beta}+i\Xi_{\alpha}\varphi_{\alpha} 
\notag \\
& -\frac{1}{12}\Psi_{\alpha\beta}C_{\alpha\beta\lambda\delta}\varphi
_{\lambda}\varphi_{\delta}\}  \notag
\end{align}

\vspace{0in}

\vspace{0in}The integral over $\varphi $ is Gaussian and can be done
exactly. \hspace{0pt}If we set 
\begin{equation}
A_{\alpha \beta }=V_{\alpha \beta }^{-1}+C_{\alpha \beta }+\frac{1}{6}\Psi
_{\lambda \delta }C_{\alpha \beta \lambda \delta }  \tag{B-2}
\end{equation}
we have: 
\begin{equation}
\begin{array}{cc}
\mathcal{Z}= & \int \mathcal{D\Phi } \mathcal{D}\Psi \exp \{\frac{-1}{4!}%
\Psi _{\alpha \beta }C_{\alpha \beta \lambda \gamma }\Psi _{\lambda \gamma }-%
\frac{1}{2}\rho _{\alpha }^{0}A_{\alpha \beta }^{-1}\rho _{\beta }^{0}-\frac{%
1}{2}\text{Tr}\log A_{\alpha \beta }\} \\ 
& \times \int \mathcal{D}\mu \exp \{-\frac{1}{2}\mu _{\alpha }V_{\alpha
\beta }^{-1}\mu _{\beta }-i\mu _{\alpha }\Phi _{\alpha }-i\rho _{\alpha
}^{0}\mu _{\alpha }+\frac{1}{2}\mu _{\alpha }V_{\alpha \beta }^{-1}A_{\beta
\lambda }^{-1}V_{\lambda \gamma }^{-1}\mu _{\gamma } \\ 
& +i\mu _{\alpha }V_{\alpha \beta }^{-1}A_{\beta \lambda }^{-1}\rho
_{\lambda }^{0}\}
\end{array}
\tag{B-3}
\end{equation}
Again the $\mu -$integral is Gaussian. After integrating out $\mu ,$ we have 
\begin{align}
\mathcal{Z}& =\int \mathcal{D\Phi D}\Psi \exp \{\frac{-1}{4!}\Psi _{\alpha
\beta }C_{\alpha \beta \lambda \gamma }\Psi _{\lambda \gamma }  \tag{B-4} \\
& -\frac{1}{2}\rho _{\alpha }^{0}A_{\alpha \beta }^{-1}\rho _{\beta }^{0}-%
\frac{1}{2}\text{Tr}\log A_{\alpha \beta }  \notag \\
& -\frac{1}{2}\zeta _{\alpha }B_{\alpha \beta }^{-1}\zeta _{\beta }-\frac{1}{%
2}\text{Tr}\log B_{\alpha \beta }  \notag \\
& :=\int \mathcal{D\Phi D}\Psi \exp \{-\mathcal{F}(\Phi ,\Psi )\}  \notag
\end{align}

\vspace{0in}If we set 
\begin{equation}
T^{-1}=V^{-1}-V^{-1}UV^{-1} , \tag{B-5}
\end{equation}
leave out terms of order ($C_{\alpha \beta \lambda \gamma })^{2},$
and use the incompressibility constraint, i.e., 
\begin{align}
\rho _{2}^{0}& =0  \tag{B-6} \\
\Phi _{1}& =0  \notag
\end{align}

\vspace{0in}we get after some quite heavy algebra the following simple form: 
\begin{align}
\mathcal{F}(\Phi,\Lambda) & =\frac{1}{2}\int dx_{1}dx_{2}\Phi(1)T(1,2)\Phi
(2)  \tag{B-7} \\
& -\frac{1}{4!}\int dx_{1}...dx_{4}\Lambda(1,2)C_{2222}^{-1}(1,2,3,4)\Lambda
(3,4)  \notag \\
& -\frac{1}{2}\int dx_{1}...dx_{8}\Phi(1)(TV^{-1}U)(1,3)\Lambda
(4,5)(UV^{-1}T)(5,8)\Phi(8)  \notag \\
& +\frac{1}{12}\int dx_{1}dx_{2}\Lambda(1,2)U(2,1)  \notag \\
& +\frac{1}{12}\int dx_{1}...dx_{6}(TV^{-1}U)(1,4)\Lambda(4,5)(UV^{-1}T)(6,1)
\notag
\end{align}
where 
\begin{equation}
\Lambda(x_{1},x_{2})=\int dx_{3}dx_{4}C_{2222}(1,2,3,4)\Psi(3,4)  \tag{B-8}
\end{equation}

and 
\begin{equation*}
\Phi(x)=\Phi_{2}(x)
\end{equation*}

Now we can integrate over $\Lambda $ since the integral is only Gaussian,
and we find that 
\begin{equation}
\mathcal{Z}=\int \mathcal{D\Phi }\exp \{-\mathcal{F}(\Phi )\}  \tag{B-9}
\end{equation}
where 
\begin{align}
\Delta \mathcal{F}(\Phi )& =\mathcal{F}(\Phi )-\mathcal{F}_{0}  \tag{B-10}
\\
& =\frac{1}{2 vol}\sum_{q}\Phi (q) \lbrack T(q)  \notag \\
& -\frac{1}{6}\frac{1}{vol}\sum_{p}\frac{C_{2222}(q,-q,p,-p)}{%
[C_{22}(q)]^{2}C_{22}(p)} \rbrack \Phi (-q)  \notag \\
& -\frac{1}{24}\frac{1}{(vol)^{3}}\sum_{qpk}\frac{C_{2222}(q,p,k,-(q+p+k))}{%
C_{22}(q)C_{22}(p)C_{22}(k)C_{22}(q+p+k)}  \notag \\
& \times \Phi (q)\Phi (p)\Phi (k)\Phi (-p-q-k)  \notag
\end{align}
and 
\begin{equation}
T(q)=\frac{1}{C_{22}(q)}+V_{22}(q) \tag{B-11}
\end{equation}
is now the effective potential for the two-component incompressible copolymer
melt.

%% ....................................

\end{document}